\documentclass[11pt]{article}

\renewcommand{\arraystretch}{1.3}
\makeatletter
\newdimen\normalarrayskip              
\newdimen\minarrayskip                 
\normalarrayskip\baselineskip \minarrayskip\jot
\newif\ifold             \oldtrue            \def\new{\oldfalse}
\def\arraymode{\ifold\relax\else\displaystyle\fi} 
\def\eqnumphantom{\phantom{(\theequation)}}     
\def\@arrayskip{\ifold\baselineskip\z@\lineskip\z@
     \else
     \baselineskip\minarrayskip\lineskip2\minarrayskip\fi}
\def\@arrayclassz{\ifcase \@lastchclass \@acolampacol \or
\@ampacol \or \or \or \@addamp \or
   \@acolampacol \or \@firstampfalse \@acol \fi
\edef\@preamble{\@preamble
  \ifcase \@chnum
     \hfil$\relax\arraymode\@sharp$\hfil
     \or $\relax\arraymode\@sharp$\hfil
     \or \hfil$\relax\arraymode\@sharp$\fi}}
\def\@array[#1]#2{\setbox\@arstrutbox=\hbox{\vrule
     height\arraystretch \ht\strutbox
     depth\arraystretch \dp\strutbox
     width\z@}\@mkpream{#2}\edef\@preamble{\halign
\noexpand\@halignto
\bgroup \tabskip\z@ \@arstrut \@preamble \tabskip\z@ \cr}%
\let\@startpbox\@@startpbox \let\@endpbox\@@endpbox
  \if #1t\vtop \else \if#1b\vbox \else \vcenter \fi\fi
  \bgroup \let\par\relax
  \let\@sharp##\let\protect\relax
  \@arrayskip\@preamble}
\def\eqnarray{\stepcounter{equation}%
              \let\@currentlabel=\theequation
              \global\@eqnswtrue
              \global\@eqcnt\z@
              \tabskip\@centering
              \let\\=\@eqncr
 \halign to \displaywidth\bgroup
    \eqnumphantom\@eqnsel\hskip\@centering
    $\displaystyle \tabskip\z@ {##}$%
    \global\@eqcnt\@ne \hskip 2\arraycolsep
         $\displaystyle\arraymode{##}$\hfil
    \global\@eqcnt\tw@ \hskip 2\arraycolsep
         $\displaystyle\tabskip\z@{##}$\hfil
         \tabskip\@centering
    &{##}\tabskip\z@\cr}
\begingroup\ifx\undefined\newsymbol \else\def\input#1 {\endgroup}\fi

\catcode`\@=11
\def\marginnote#1{}

\newcount\hour
\newcount\minute
\newtoks\amorpm
\hour=\time\divide\hour by60 \minute=\time{\multiply\hour by60
\global\advance\minute by-\hour}
\edef\standardtime{{\ifnum\hour<12 \global\amorpm={am}%
        \else\global\amorpm={pm}\advance\hour by-12 \fi
        \ifnum\hour=0 \hour=12 \fi
        \number\hour:\ifnum\minute<10 0\fi\number\minute\the\amorpm}}
\edef\militarytime{\number\hour:\ifnum\minute<10 0\fi\number\minute}

\def\draftlabel#1{{\@bsphack\if@filesw {\let\thepage\relax
      \xdef\@gtempa{\write\@auxout{\string
          \newlabel{#1}{{\@currentlabel}{\thepage}}}}}\@gtempa \if@nobreak
    \ifvmode\nobreak\fi\fi\fi\@esphack} \gdef\@eqnlabel{#1}}
    \def\@eqnlabel{}
\def\@vacuum{}
\def\draftmarginnote#1{\marginpar{\raggedright\scriptsize\tt#1}}

\def\draft{
  \oddsidemargin -.5truein
  \def\@oddfoot{\footnotesize \sl preliminary draft \hfil
    \rm\thepage\hfil\sl\today\quad\militarytime}
  \let\@evenfoot\@oddfoot \overfullrule 3pt
    \let\label=\draftlabel
    \let\marginnote=\draftmarginnote
  \def\@eqnnum{(\theequation)\rlap{\kern\marginparsep\tt\@eqnlabel}%
    \global\let\@eqnlabel\@vacuum}

  }

\def\be{\begin{eqnarray}}
\def\ee{\end{eqnarray}}
\def\nn{\nonumber}

\def\p{\partial}

\def\beq{\begin{equation}}
\def\eeq{\end{equation}}
\def\ba{\beq\new\begin{array}{c}}
\def\ea{\end{array}\eeq}
\def\be{\ba}
\def\ee{\ea}

\newfont{\alef}{msbm10 at 12pt}
\newfont {\goth}{eufm10 at 11pt}
\def\mathbb#1{\hbox{{\alef #1}}}

\let\@@savethanks\thanks
\def\thanks#1{\gdef\thefootnote{\alph{footnote}}\@@savethanks{#1}}

\title{{\bf M-Theory of Matrix Models
} \vspace{.5cm}}
\author{{\bf A.Alexandrov}\thanks{E-mail: \ al@itep.ru}
\date{ } \\ {\small
{\it ITEP, Moscow, Russia}}\\ \\
{\bf A. Mironov}\footnote{E-mail: \ mironov@itep.ru; mironov@lpi.ru}
\date{ } \\
{\small {\it Lebedev Physics Institute}
and {\it ITEP, Moscow, Russia}}\\ \\
{\bf A.Morozov}\thanks{E-mail: \ morozov@itep.ru}
\date{ } \\ {\small {\it ITEP, Moscow, Russia}}
}

\begin{document}

\maketitle

\vspace{-10.5cm}

\begin{center}
\hfill FIAN/TD-3/06\\
\hfill ITEP/TH-20/06\\
\end{center}

\vspace{9.0cm}

\begin{abstract}
\noindent Small $M$-theories unify various models of a given family
in the same way as the $M$-theory unifies a variety of superstring
models. We consider this idea in application to the family of
eigenvalue matrix models: their $M$-theory unifies various branches
of Hermitean matrix model (including Dijkgraaf-Vafa partition
functions) with Kontsevich $\tau$-function. Moreover, the
corresponding duality relations look like direct analogues of
instanton and meron decompositions, familiar from Yang-Mills theory.
\end{abstract}

\bigskip

\def\thefootnote{\arabic{footnote}}

\section{Introduction}

String theory\footnote{
We use this term in the spirit of ref.\cite{UFN2}, as a name
for the broad modern version of quantum field theory (QFT) with
numerous applications, of which various superstring models
of sub-Planckian physics are important, but particular examples.}
approach to analysis of QFT models involves the following seven
steps.

\bigskip

1. When one introduces particular QFT model it is usually defined as
a perturbation expansion around a given {``classical"} background.

2. Perturbation theory is usually defined in terms of correlators
that are {\it Gaussian} integrals. This leads to two corollaries.
The nice one is the Wick theorem and the powerful diagram technique.
The bad one is divergence of the perturbation series, which is
asymptotic series due to the standard Dyson argument. This is causes
by expanding integrands without respect to their asymptotics and,
therefore, non-trivial phase/branch structure.\footnote{Some more
clever non-Gaussian expansions where weights are dictated by
dominant non-quadratic terms in action are already in agenda (see
\cite{cubin,DM2} for different conceptual approaches to the
problem), but are still far from any practical significance. }

3. The divergent perturbation series can be often presented in the
form of a ``genus expansion" asymptotic series in a single
``Planck-constant" parameter, particular terms of the genus
expansion being {\it convergent} multiple series in numerous other
perturbation parameters (called coupling constants or
time-variables).

4. Non-perturbative partition function gets contributions from a
variety of ``classical" backgrounds, each contribution dressed by
its own perturbative expansion.

\bigskip

At this stage we get a representation of non-perturbative partition
function -- i.e. the only quantity which has an objective physical
meaning -- as a peculiar {\it triple} expansion: over ``classical
backgrounds",  asymptotical quasiclassical (genus) expansion near
each background, convergent series in powers of coupling constants
for each term of the genus expansion. Actually the boundaries
between the three levels of hierarchy are not well defined: changes
of integration variables easily mix the levels. Worse than that,
given this triple expansion (of which we can usually calculate only
the first terms), it is rather difficult to understand the structure
of the globally defined non-perturbative partition function (beyond
any doubt, it is a nice function, just with sophisticated analytical
structure, with numerous branchings and, often, essential
singularities, which make perturbation series numerous, divergent
and difficult to sew). The next three steps are targeted at curing
this situation.

\bigskip

5. Of the three levels the highest -- sum over backgrounds -- is
understood the worst. It is often formulated in the language of
``instantons": the idea is that this sum has its own hierarchical
structure and infinitely many terms in the sum are somehow expressed
through multi-instanton configurations (that probably can be build
from the single-instanton ones by acting a few operator generators).
A nice realizations of this idea exist at the level of ordinary
quantum mechanics see \cite{ABC,Put} and in $4d$ QFT in the
supersymmetric context (where perturbative contributions
artificially cancel) \cite{Nek}.

6. It is well known, however, that instanton calculus is hardly
exhaustive: the proper generators, if any, would rather be made from
instanton elementary constituents -- {\it merons}
\cite{CDG,MMT}.\footnote{ For Yang-Mills theory it is a puzzling
problem: Yang-Mills instanton can be described as a process of
creation and annihilation of a monopole-antimonopole pair, splitting
of instanton into two merons is used to separate contributions of
these individual quasiparticles and can provide proper variables to
describe monopole condensation. Thus, the meron calculus should play
the central role in confinement mechanism \cite{CDG}. At the same
time there is still no clear room for merons in the ADHM
construction \cite{ADHM}. } Also the role of merons in the
distribution of roles between the sum over backgrounds and genus
expansions, i.e. between the two upper levels of triple expansion,
remains under-investigated (as almost everything in meron physics).

7. Different kinds of expansions -- emerging when one starts from
different QFT models at step 1 -- at non-perturbative level are
unified in a global partition function. Since it unifies {\it a
priori} different models it deserves a special name: nowadays it is
often called $M$-theory. Being a true result of {\it functional
integration}, the global partition function inherits a hidden
symmetry \cite{UFN3} (coming from the freedom to change integration
variables and providing a non-trivial generalization of the
Ward-like symmetries): it belongs to a narrow class of integrable
$\tau$-functions \cite{tauf}, satisfying physically-mysterious
quadratic relations \cite{UFN3,quadr}. Various models, arising in
different limits in the space of couplings (when one or another
background gives dominant contribution), are said to be related by
{\it duality}: sometime, but not always, duality transformations
form a small group (but not obligatory $Z_2$ or $SL(2)$, some
dualities can look like {\it trialities} etc, generically there is
no small group at all). Switching on more and more perturbations,
one enlarges $M$-theory, and at the end of the day one can arrive at
the all-unifying ``theory of everything" (nicknamed {\it String
Theory} in \cite{UFN2}), where every two QFT models are ``dual" in
the sense that complicated correlators evaluated in one of them are
converted by an appropriate analytical continuation into other
complicated correlators in another one. However, one can stop at
intermediate stages: by restricting classes of correlators (for
example, to those of polynomials of original fields ), one obtains
$M$-theories for sufficiently small varieties of QFT models.

\bigskip

The task of the present paper will be a brief introduction into
$M$-theory of ordinary ({\it eigenvalue}) matrix models \cite{UFN3}.
This presentation is based on our recent paper \cite{ammMth}, which
provided a unified derivation of significant results about relations
between {\it a priori} different matrix models (Kontsevich model
\cite{Ko}, complex-matrix model \cite{coma,MMMM} and Hermitean model
\cite{hemo,Tch} with its many phases \cite{hemophases}, including
the Dijkgraaf-Vafa \cite{DV} and other \cite{ammChops} branches),
obtained during the last decade \cite{Ch1}-\cite{Giv} and based, in
turn, on the long studies of eigenvalue models, since they were
first introduced by F.Dyson in 1962 \cite{hemo}.


The duality relations
connect the partition
functions of the Hermitean and Kontsevich matrix models. The third
important eigenvalue model in the same $M$-theory is less popular
but still important {\it complex} matrix model with partition
function $Z_{C}(t)$. Actually, duality expresses $Z_{C}$ through
$Z_K$ \cite{ammMth}.

Like the Hermitean model, the complex one has many phases/branches,
differing by shifts of the time-variables, like (13) and (15). We
reserve the notation $Z_C(t)$ for the simplest, Gaussian phase
(similarly, in the present paper $Z_K(\tau)$ denotes the simplest,
Kontsevich branch of the GKM partition function; for dualities
between different branches of GKM see \cite{KM}).

Other models,
especially the very important unitary one \cite{umo}, can probably
be also included into this matrix-model $M$-theory.\footnote{To
avoid a confusion, we remind that the term ``M-theory" was
introduced in \cite{Mth} for a unified description of the web of
superstring models.
Progress with this superstring $M$-theory, revived interest
\cite{landsc} to the old landscape-style views on string theory
(e.g. in \cite{GLM,UFN2}) and finally brought more attention to the
string theory {\it per se}, irrespective of its particular
applications -- even that great as unification of all fundamental
interactions and of all cosmological scenarios. The $M$-theory of
matrix models, considered in the present paper, has no direct
relation to the superstring $M$-theory (which also uses the
matrix-models technique \cite{BS}). Though mathematically and
conceptually similar (and this is what makes our study important),
they concern absolutely different physical theories, the very small
one, even with no time-evolution (but allowing a detailed and
rigorous description), and enormously big one, big enough to include
all our world along with entire variety of all other worlds,
thinkable and unthinkable. }

\section{The WEB of matrix-model dualities}

\subsection{Basic instanton and meron decompositions}

Partition function of the Hermitean matrix model has many branches,
some are labeled by two functions $W(\xi)$ and $f(\xi)$
\cite{hemophases,ammChops}: $Z_{W,f}(t)$. These phases are
distinguished by existence of a special -- t'Hooft's -- genus
expansion, where genus is indeed the genus of the Feynman-t'Hooft
fat-graph diagrams. The leading (spherical) terms of this expansion
are known as Dijkgraaf-Vafa partition functions \cite{DV}. The best
studied is the Gaussian phase, $Z_G(t) =
Z_{\frac{1}{2}M\xi^2,4MS}(t)$, associated with $W(\xi) =
\frac{1}{2}M\xi^2$ and $f(\xi) = 4MS = const$. Here $S=Ng$, $N$ being the
size of the Hermitean matrix. $Z_G(t)$ is the
Toda-chain $\tau$-function \cite{Tch}, while other $Z_{W,f}(t)$ are
$\tau$-functions of less known hierarchies.

For polynomial $W$ and $f$, $Z_{W,f}$ can be expressed through the
Gaussian $Z_G(t)$, like the multi-instanton partition function would
be expressed through the single-instanton one. If the derivative
$W'(\xi)$ is a polynomial of degree $n$ in $\xi$, $W'(\xi) =
\prod_{i=1}^n (\xi-\xi_i)$, \be Z_{W,f}(t) = \hat
U_{W|G}\big(t|t^{(1)},\ldots t^{(n)}\big) \Big\{\otimes_{i=1}^n
Z_G^{(i)}(t^{(i)}) \Big\} \label{W|G} \ee where time-variables
$t^{(i)}$ are obtained from $t$ by shifts, so that they describe
expansion around the extremum $\xi_i$ of $W(\xi)$, and parameters
$M_i$ and $N_i$ of $Z_G^{(i)}$ are made from $W''(\xi_i)$ and
$f(\xi_i)$, see \cite{David,Th,Giv,hemophases,ammChops} for the
details. Operator $\hat U\big(t|t^{(1)},\ldots t^{(n)}\big)$
intertwines between elementary constituents, such intertwiners will
appear in all duality relations. All these operators have a very
special form, typical for the Weyl-Moyal operators \cite{Moyal,GLM}
and their Batalin-Vilkovisky generalizations \cite{BV}: they are
quadratic exponentials of the time-variables $t$ and/or
time-derivatives.

However, the decomposition formula (\ref{W|G}) is only the
beginning. Like Yang-Mills instanton is made from two merons, the
Gaussian partition function $Z_G(t)$ decomposes into the two
Kontsevich $\tau$-functions $Z_K(\tau)$ \cite{ammMth}: \be Z_G(t) =
\hat U_{G|K}(t|\tau_\pm) \Big\{ Z_K(\tau_+)\otimes Z_K(\tau_-)\Big\}
\label{G|K} \ee where the variables $\tau_\pm$ can be expressed
through $t$ in a variety of ways (for particular $t-\tau_\pm$
relations this formula was first suggested in \cite{MMMM} and
carefully derived in \cite{Ch1}). Accordingly the meron
decomposition exists for the other branches described by the
polynomial $W$ of degree $n+1$: \be Z_{W,f}(t) = \hat
U_{W|K}\big(t|\tau^{(1)}_\pm,\ldots,\tau^{(n)}_\pm\big)
\Big\{\otimes_{i=1}^n \left(Z_K^{(i)}(\tau^{(i)}_+) \otimes
Z_K^{(i)}(\tau^{(i)}_-\right)\Big\} \label{W|K} \ee However, in
variance with (\ref{G|K}), this formula is not fully investigated in
\cite{ammMth}, operator $\hat U_{W|K}$ is not yet explicitly
constructed and the relation \be \hat U_{W|G} \otimes_{i=1}^n \hat
U^{(i)}_{G|K} = \hat U_{W|K} \ee also requires direct derivation.
Entire theory of intertwining Weyl-Moyal operators and underlying
$*$-structures \cite{ammMth} remains to be built.

\subsection{Polynomial representations (moment variables)}

$Z_G(t)$ and $Z_K(\tau)$ are themselves invariant under some
restricted changes of variables \cite{ammMth}: \be Z_G(t) =
U_{G|G}(t|\tilde t) Z_G(\tilde t) \label{G|G} \ee and \be Z_K(\tau)
= U_{K|K}(\tau |\tilde\tau) Z_K(\tilde\tau) \label{K|K} \ee In these
formulas intertwiners $U_{G|G}$ and $U_{K|K}$ do not contain
time-derivatives: they are just functions of time-variables and we
do not write {\it hats} over them. As in all previous cases we do
not write the somewhat lengthy formulas for transformations of time
variables explicitly (they can be found in ref.\cite{ammMth}). In
(\ref{G|G}) and (\ref{K|K}) these transformations are linear and
depend on two free parameters. However, these few parameters can
themselves be made {\it arbitrary} functions of times, and the
two-parametric families of {\it linear} transformations generate
infinitely large (still special) families of {\it non-linear}
transformations. Moreover, among these non-linear transformations
there is one that greatly simplifies $Z_G(t)$ and $Z_K(\tau)$:
namely, all except the first two terms in genus expansions can be
made {\it polynomial} in time-variables. We call these special
choices $\bar t$ and $\bar\tau$, and actually \be \bar Z_G(t) = \bar
Z_G(\bar t) \label{G|Gbar} \ee and \be \bar Z_K(\tau) = \bar
Z_K(\bar\tau) \label{K|Kbar} \ee In these formulas $\bar Z$ is
defined in the following way: if $\log Z = \sum_{p\geq 0}
g^{2p-2}F_p$ is the genus expansion for $Z$, then $\log \bar Z =
\sum_{p\geq 2} g^{2p-2}\bar F_p$ and, all $\bar F_p$ with $p\geq 2$
are {\it polynomials} of $\bar t$ or $\bar\tau$ variables (with
powers growing as $p$ increases). Since in (\ref{G|G}) and
(\ref{K|K}) intertwiners are {\it functions}, and -- as a corollary
-- in this case contain only terms with $g^{-2}$ and $g^0$ in the
exponent, they do not show up in (\ref{G|Gbar}) and (\ref{K|Kbar}).
These formulas are the best possible illustration of statement $3$
in the Introduction: contributions of given genus are not just
convergent, in appropriate coordinates they are polynomial.

The only word of caution to be made is that while (\ref{K|Kbar})
holds for conventional (t'Hooft's) genus expansion, in
(\ref{G|Gbar}) another, -- {\it loop} -- expansion is implied, see
\cite{ammMth}. The polynomial coordinates exist the in Gaussian case
for the t'Hooft expansion as well \cite{ACKM}, associated variables
are known as {\it moments}, but unlike (\ref{G|Gbar}) the
corresponding relation \be Z_G(t) = \hat U_{G|ACKM}(t|\mu)
Z_{ACKM}(\mu) \label{G|ACM} \ee involves a new partition function
$Z_{ACKM}(\mu)$: contributions of genera $p\geq 2$ to $\log
Z_{ACKM}(\mu)$ are polynomials in $\mu$, but $Z_{ACKM}$ and $Z_G$
are different functions.\footnote{To avoid possible confusion, we
give a trivial example illustrating the difference between
(\ref{G|G}) and (\ref{G|ACM}):
$$ \sin t = \sin \bar t = \frac{1}{2i}\big(\mu - \mu^{-1}\big)$$
where $\bar t = t + 2\pi$ and $\mu = e^{it}$. The first equality is
a model of (\ref{G|G}) with a single function $Z(t) = \sin(t)$,
while the second equality relates two {\it different} functions,
$\sin x$ and $\frac{1}{2i}(x-x^{-1})$. }

\section{Group theory approach}

Partition functions of eigenvalue matrix models are best defined as
the highest weight vectors in the special Verma modules of Virasoro
algebras,
i.e. are 
eigenvectors of their maximal (Borel) subgroups: \be \hat L^I_- Z_I
= 0 \ee Here $I = G,K,C$ and Virasoro loop operators are: \be \hat
L^G_-(z) = \sum_{n=-1}^{\infty} \frac{dz^2}{z^{n+2}}\left\{
\sum_{k=0}^\infty kt'_k\frac{\partial}{\partial t_{k+n}} + g^2
\sum_{a+b=n} \frac{\partial^2}{\partial t_a\partial t_b}\right\},
\nn\\
\hat L^K_-(z) = \sum_{n=-1}^{\infty} \frac{dz^2}{z^{n+2}}\left\{
\sum_{k=0}^\infty \big(k+1/2\big)\tau'_k
\frac{\partial}{\partial \tau_{k+n}} + g^2
\sum_{a+b=n-1} \frac{\partial^2}
{\partial \tau_a\partial \tau_b}\right\} +
\frac{1}{16g^2}\tau_0^2\delta_{n,-1} + \frac{1}{16}\delta_{n,0},
\nn\\
\hat L^C_-(z) = \sum_{n=0}^{\infty} \frac{dz^2}{z^{n+2}}\left\{
\sum_{k=0}^\infty kt'_{2k}\frac{\partial}{\partial t_{2k+2n}} +
g^2\sum_{a+b=n}\frac{\partial^2} {\partial t_{2a}\partial
t_{2b}}\right\} \ee In these formulas \be t'_k =
-\frac{1}{2}M\delta_{k,2} + t_k, \ \ \ \tau_k' =
-\frac{2}{3}M\delta_{k,1} + \tau_k \ee $Z_{W,f}$ satisfies the
equations \be \hat L^W_- Z_{W,f} = 0 \ee where $\hat L^W_-(z)$
differs from $\hat L^G_-(z)$ only by a shift: for generic $W(\xi)$
it can be written as \be \sum_k kt'_k\xi^{k-1} = -W'(\xi) + \sum_k
kt_k\xi^{k-1} \ee The function $f(\xi)$ parameterizes different
formal-series solutions to this loop equation, possessing t'Hooft's
genus expansion. In the Gaussian case of $W(\xi) =
\frac{1}{2}M\xi^2$, it reduces to $f(\xi) = 4MS = {\rm const}$.

It is expected that similar representations exist for generic
non-perturbative partition functions (though even the proper
analogues of Virasoro algebras are not yet known in most cases).
Thus, the relevant formalism of {\it non-linear} realizations of
symmetries is of universal importance. The realizations are
non-linear because of the {\it shifts} of time-variables in the
Virasoro equations. To understand the structure and symmetries
(dualities) of eigenvectors in non-linear representations, it
deserves to begin from the simplest examples, with ordinary
operators instead of the loop ones. We return to matrix models in
section \ref{webdu}.

\section{Examples of non-linear realizations of symmetries}

In this section $x,y,z,t,\tau$ will play the role of the
time-variables $t$ and $\tau$ of the previous sections.

\subsection{$\hbox{\alef R}^*\otimes SO(2)$ group}

Consider the $SO(2)$ invariant functions $Z_N$ on the real plane
$\hbox{\alef R}^2$ that realize the weight $N$ representation of the
group of multiplications by real numbers, \be\label{1}
\left(y'\frac{\partial}{\partial x} - x'\frac{\partial}{\partial
y}\right) Z_N = 0 \\ \left(x'\frac{\partial}{\partial x} +
y'\frac{\partial}{\partial y}\right) Z_N = NZ \ee where the shifted
variables $x'=R\cos\phi =x+r$, $y'=R\sin\phi =y$ parameterize the
real plane. Since there is no central extension, the maximal
subalgebra in this case can be taken to coincide with the entire
algebra.

Now one may use the two different strategies to find $Z_N$. First of
all, one can just note that the $SO(2)$ invariant functions on the
real plane $\hbox{\alef R}^2$ depend only on the radial variable
$R$, while the second equation in (\ref{1}), the homogeneity
condition implies that $Z_N\sim R^N$. The main subtlety here is
related to the shift of coordinates $(x',y')\to(x,y)$ so that one
should re-calculate the linearly realized $SO(2)$ symmetry in
coordinates $(x',y')$ ($\phi \rightarrow \phi + \alpha$) to its
non-linear realization in coordinates with shifted origin $(x,y)$.

Thus the ``partition function" \be Z_N = {\rm const}\cdot R^N \sim
\left(1 + \frac{2x}{r} + \frac{x^2+y^2}{r^2}\right)^{N/2}
\label{rotZdef} \ee

The other possible way of calculating is to construct the solution
to (\ref{1}) as a power series in $x$ and $y$, normalized to start
with unity:
\be Z_N(x,y) = \left(1 + \frac{2x}{r} + \frac{x^2+y^2}{r^2}
\right)^{N/2} = 1 + \frac{N}{r}x + \frac{N(N-1)}{2r^2}x^2 +
\frac{N}{2r^2}y^2 + \ldots \ee Coefficients in this series can be
recurrently determined from (\ref{1}) (this is the only way to
construct solutions available in matrix models).

To compare these two approaches, one can note that there is the
identity: \be \forall \alpha \ \ \ Z(x,y) = Z\Big(-r(1-\cos\alpha) +
x\cos\alpha + y\sin\alpha,\ -r\sin\alpha-x\sin\alpha +
y\cos\alpha\Big) \ee Let us check it with for the power series: \be
1 + \frac{N}{r}\Big(-r(1-\cos\alpha) + x\cos\alpha +
y\sin\alpha\Big) + \frac{N(N-1)}{2r^2}\Big(-r(1-\cos\alpha) + \nn \\
+ x\cos\alpha + y\sin\alpha\Big)^2 +
\frac{N}{2r^2}\Big(-r\sin\alpha-x\sin\alpha + y\cos\alpha\Big)^2 +
\ldots = \nn \\ = 1 +  \frac{N}{r}x + \frac{N(N-1)}{2r^2}x^2 +
\frac{N}{2r^2}y^2 + \ldots \label{rotZ1} \ee where both sides should
be understood as power series in $x$,$y$ and $\alpha$. For example,
the terms without $x$ and $y$ are \be -N(1-\cos\alpha) +
\frac{1}{2}N\sin^2\alpha + \frac{1}{2}N(N-1)(1-\cos\alpha)^2 +
\ldots = \nn \\ = \frac{N}{2}\left(\sin^2\alpha -
4\sin^2\frac{\alpha}{2} -4\sin^4\frac{\alpha}{2} + \ldots \right) +
\frac{N^2}{2}\left(4\sin^4\frac{\alpha}{2} + \ldots \right) + \ldots
\ee For $\alpha = \frac{\pi}{2}$ one gets \be Z_N(x,y) = Z_N(-r +
y,\ -r-x) \label{rotZ} \ee This is trivially true, if one knows that
$Z$ is a function of $(r+x)^2+y^2$, but is a non-trivial identity
for formal series, if one does not know their sums explicitly.

\subsection{$T$-duality}

Much closer to the case of matrix models is the following
example\footnote{A simple one-dimensional example that is very close
to the matrix model case is considered in detail in \cite{Mir}.}:
\be \left(t\frac{\partial}{\partial t} + \frac{\partial^2}{\partial
x^2}\right) Z = 0 \label{toyvir} \ee By a change of variable $t =
e^{i\tau}$ it is transformed into the heat/Shr\"odinger equation
(among other things, this $t-\tau$ relation emphasizes
``un-naturalness" of expansions in integer powers of $t$, the
expansion around the point $t=0$ is not very nice from the point of
view of the heat equation): \be \left(-i\frac{\partial}{\partial
\tau} + \frac{\partial^2}{\partial x^2}\right) Z = 0 \ee with the
generic solution \be Z(x|\tau) = \int e^{ipx} e^{ip^2\tau}c(p)dp =
\int t^{p^2}e^{ipx} c(p)dp \label{Zvsp} \ee In this $2$-variable
example it is obvious that (\ref{toyvir}) defines $Z(x|\tau)$
ambiguously, and solutions are parameterized by a function $c(p)$: a
counterpart of $f(\xi)$ in $Z_{W,f}$.

Particular solutions of the heat equation include:

$\bullet$
the heat kernel on a complex line = real plane
\be
Z_1(x|\tau) = \sqrt{\frac{i\pi}{\tau}} e^{-ix^2/4\tau}, \ \ \
{\rm with} \ \ c_1(p) =1;
\ee

$\bullet$ the periodic Jacobi $\theta$-function (actually, this
$\theta = \theta_{00}$, but we omit the index $00$ in what follows)
\be Z_2(x|\tau) =
\theta\left(\frac{x}{2\pi}\left|\frac{\tau}{\pi}\right)\right.  =
\sum_{n=-\infty}^\infty e^{in^2\tau}e^{inx},\ \ \ {\rm with} \ \
c_2(p) = \sum_{n=-\infty}^\infty \delta(p-n); \ee

$\bullet$ the heat kernel on a torus \be Z_3(x|\tau) =
\sqrt{\frac{{i\pi}}{{\tau}}}\ e^{-i x^2/4\tau}
\theta\Big(\frac{x}{2\tau}\left|-\frac{\pi}{\tau}\Big)\right. =
\sqrt{\frac{{i\pi}}{{\tau}}}\ e^{-i x^2/4\tau}
\sum_{n=-\infty}^\infty e^{-i\pi^2 n^2/\tau}e^{i\pi nx/\tau}. \ee
Function $c_3(p)$, associated with $Z_3$, actually coincides with
$c_2(p)$: this is the corollary of $T$-duality, which states that
\be Z_2(x|\tau) = Z_3(x|\tau) \ee

Usually $T$-duality is treated as a {\it transcendental} relation,
derived, for example, with the help of Poisson re-summation trick.
However, one may represent {\it all} solutions to the heat equation,
including $Z_1$, $Z_2$ and $Z_3$, as linear combinations of Hermite
rescaled polynomials ${\rm He}_n(x|\tau) \equiv
\left(-i\tau\right)^{n/2} H_n\left(\displaystyle{ix\over
2\sqrt{-i\tau}}\right)$ (where we denoted through $H_n$ the standard
Hermite polynomials)
$$
{\rm He}_0(x|\tau) =1,\ \ \ \
{\rm He}_1(x|\tau) =ix,\ \  \ \
{\rm He}_2(x|\tau) =2i\tau - x^2,\ \ \ \
{\rm He}_3(x|\tau) =-6x\tau - ix^3,
$$ $$
{\rm He}_4(x|\tau) =-12\tau^2 - 12ix^2\tau + x^4, \ \ \ldots
$$
since they solve the heat equation. Then, the relations between
different partition functions become involved but {\it elementary}
algebraic relations.

For instance, in representation (\ref{Zvsp}) \be e^{ip^2\tau}e^{ipx}
= \sum_{n=0}^\infty \frac{p^n}{n!} {\rm He}_n(x|\tau) \ee while, in
order to get an expansion of the dual representation into the
Hermite polynomials, one needs to shift the variable $\tau
\rightarrow \tau' = T + \tau$ and expand in inverse powers of the
background $T$. First, we expand in this way $Z_1$:\footnote{This
formula is a particular case of the celebrated M\"oller formula,
\cite[eq.(10.13.22)]{BE}\be {1\over\sqrt{1-z^2}}\
\exp\left({2xyz-(x^2+y^2)z^2\over 1-z^2}\right)=\sum_{n=0}^\infty
{1\over n!}\left({z\over 2}\right)^nH_n(x)H_n(y)\ee} \be
\frac{1}{\sqrt{\tau}}e^{-ix^2/4\tau} \longrightarrow
\frac{1}{\sqrt{T}}\Big(1+\frac{\tau}{T}\Big)^{-1/2} \exp
\left\{-i\Big(\frac{x^2}{4}\Big)\frac{1}{T}
\Big(1+\frac{\tau}{T}\Big)^{-1}\right\} = \sum_{m\geq 0}^\infty
\frac{i^m}{4^mT^{m+1/2}} \frac{{\rm He}_{2m}(x|\tau)}{m!} \ee
Now for the dual representation, one needs to shift in this formula
$x$ by $q$ to obtain\footnote{To derive this formula one suffices to
note that, for the Hermite polynomials, \be {\p H_n(x|\tau)\over\p x}=inH_{n-1}
\ee } \be
\frac{1}{\sqrt{\tau}}e^{-ix^2/4\tau} e^{iqx/\tau} e^{-iq^2/\tau}
\longrightarrow \sum_{m\geq 0}^\infty \frac{i^m}{4^mT^{m+1/2}}
\frac{{\rm He}_{2m}(x-2q|\tau)}{m!}=\sum_{k,m\geq 0}^\infty
i^{m-k}{(2q)^k\over T^{m+1/2}}{\Gamma \left(m+{1\over 2}\right)\over\Gamma
\left({1\over 2}\right)}{{\rm He}_{2m-k}(x|\tau)\over k!(2m-k)!}\ee

At the level of the heat equation, the $T$-duality states that
$$
\left(\frac{e^{-ix^2/4\tau}}{\sqrt{\tau}}\right)^{-1}
\left(-i\frac{\partial}{\partial\tau}
+ \frac{\partial^2}{\partial x^2}\right)
\frac{e^{-ix^2/4\tau}}{\sqrt{\tau}} =
-i\left.\frac{\partial}{\partial\tau}\right|_{x={\rm const}} +
\frac{\partial^2}{\partial x^2} - \frac{ix}{\tau}
\frac{\partial}{\partial x} =
$$
\be
= \frac{1}{\tau^2}
\left\{-i\left.\frac{\partial}{\partial \Big(-1/\tau\Big)}
\right|_{x/\tau = {\rm const}} +
\frac{\partial^2}{\partial\Big(x/\tau\Big)^2}\right\}
\ee
Accordingly, for every solution $Z(x|\tau)$ of the heat
equation, the $T$-dual function
\be
\tilde Z(x|\tau) = \frac{e^{-ix^2/4\tau}}{\sqrt{\tau}}
Z\Big(\frac{x}{\tau}\left|-\frac{1}{\tau}\Big)\right.
\ee
is also a solution.



\section{Spectral-surface approach \label{webdu}}

In the case of the Virasoro constraints only a non-linear
realization of the symmetry algebra is known: identification of
coordinates, where it is realized linearly, remains a puzzle. It is
known as the problem of {\it true} variables in string theory, and
it remains unresolved at all levels, from quantum gravity and entire
string theory to matrix models. Thus dualities, described by
eqs.(\ref{W|G})-(\ref{G|ACM}) at our present state of knowledge
remain non-trivial -- as would be (\ref{rotZ}) at the level of
(\ref{rotZ1}), without (\ref{rotZdef}). Therefore, these dualities
should be somehow {\it derived}. In ref.\cite{ammMth} we did it with
the help of an additional structure: by considering various loop
operators $\hat L^I_-$ as different asymptotics of a single {\it
global} loop operator defined on an auxiliary {\it spectral Riemann
surface}. The same technique is believed to be adequate for all
$M$-theories, one just needs to identify appropriate spectral
manifolds...

Definition of the global loop operator actually involves several
substructures \cite{ammMth}. Three are of universal importance, but
we formulate them in matrix-model terms, where the spectral
parameter $z$ is a single complex variable:

-- the choice of the spectral curve $\Sigma$, where the loop
parameter $z$ is taking values ($z$ does not lie just on the complex
plane, rather $z \in \Sigma$, and $\Sigma$ can be a non-trivial
complex curve, a ramified covering of the {\it bare} curve
$\Sigma_0$, which can be either $C$ or Riemann sphere $\bar C$ or
complex torus $C^*$);

-- the choice of the Krichever-Novikov-type \cite{KN} algebra, to
which the loop operators $\hat L(z)$ belong, it is specified by the
choice of the {\it structure algebra} (the Virasoro algebra in the
case of eigenvalue matrix models) and allowed singularities on a
bare spectral curve $\Sigma_0$ (in the Virasoro case allowed are
poles and square-root singularities, for $W_n$ algebras
ramifications of orders that are divisors of $n$ are allowed),
conversion to $\Sigma$ resolves the ramifications so that only poles
are allowed in $z\in \Sigma$;

-- the choice of a Borel subalgebra $\hat L_-(z)$, which can have
non-trivial eigenvectors, in terms of the spectral manifolds this
{\it minus-projection} can be represented by an integral operator,
\be \hat L_-(z) = \oint_C K(z,z')\hat L(z'). \label{minpro} \ee

Particular operators $\hat L^I_-(z)$ arise from the global $\hat
L_-(z)$ in vicinities of particular singularities, poles are
associated with $\hat L^G_-$, quadratic ramifications -- with $\hat
L^K_-$ and $\hat L^C_-$ (more general GKM $\tau$-functions would
arise if the structure algebra was $W_n$ with higher order
ramification points). Duality relations (\ref{W|G})-(\ref{W|K})
arise when the contour $C$ in (\ref{minpro}) is decomposed into
combinations of contours, surrounding different singularities. Most
analytical changes of the spectral variable $z \rightarrow \tilde
z(z)$, $\partial \tilde z/\partial\bar z = 0$ break the structure of
minus-projection, but some -- preserve it, and they give rise to
identities like (\ref{G|G})-(\ref{G|ACM}).

In the case of eigenvalue matrix models there are two additional
substructures \cite{ammMth}, which at least simplify the theory, but
can also appear to have more general significance. They reflect the
existence of free-field formalism on Riemann surfaces, and thus are
relevant to non-perturbative partition functions ($\tau$-functions)
associated with $2d$ conformal models.

-- The Virasoro algebra can be embedded into the universal
enveloping of the Kac-Moody algebra by the Sugawara construction
\cite{Suga}, and this embedding, $\hat L(z) = :\hat J(z)^2:$ can be
continued to Krichever-Novikov-type algebras (in the matrix-model
case, the simplest Kac-Moody algebra $\hat U(1)$ is relevant, but
everything should work for non-perturbative partition functions,
associated with all other Kac-Moody algebras -- this is the most
straightforward generalization of our construction);

-- An additional Seiberg-Witten structure \cite{SW,GKMMM}, a 1-form
$\Omega_{SW}(z)$ with special properties, can be used to construct
the kernel \be K(z,z') = \frac{\Omega_{SW}(z)}{\Omega_{SW}(z')}
\Big<\partial\phi(z)\ \phi(z')\Big> = \Omega_{SW}(z)k(z,z') \ee in
(\ref{minpro}) from the free-field correlator on the spectral curve;
actually this kernel can be used to define a multiplication
operation of the Kac-Moody currents \be \Big(\hat J \star \hat J\
\Big)(z) = \oint_C k(z,z'):\hat J(z')^2: \ee which plays a big role
in ``string field theory" \cite{SFT} and/or in
``background-independent" description \cite{bi} of the matrix model
$M$-theory along the lines of refs.\cite{Ey,ammEy}.

\bigskip

{\bf Acknowledgements}


\bigskip

A.A. is grateful to N.Amburg for useful discussions and V.Poberezhny for the
kind hospitality.

This work was partially supported by the Federal Program of the
Russian Ministry of Industry, Science and Technology No
40.052.1.1.1112, by the grants RFBR 03-02-17373 (Alexandrov), RFBR
04-02-16538a (Mironov), RFBR 04-02-16880 (Morozov), by the Grant of
Support for the Scientific Schools 8004.2006.2, NWO project
047.011.2004.026, INTAS project ``Current Topics in String Theory"
and ANR-05-BLAN-0029-01 project ``Geometry and Integrability in
Mathematical Physics".


\begin{thebibliography}{12}

\bibitem{UFN2} A.Morozov, Sov.Phys.Usp. (UFN) {\bf 35} (1992)
671-714.

\bibitem{cubin} J.Boh\'a\u{c}ik, and P.Pre\u{s}najder,
hep-th/0507129; hep-th/0503235.

\bibitem{DM2} V.Dolotin, and A.Morozov, {\it Introduction to Non-Linear
Algebra}, to appear; 
hep-th/0501235.

\bibitem{ABC} S.Coleman, The use of instantons, Erice lectures, 1977\\
V.Novikov, M.Shifman, A.Vainshtein, and V.Zakharov,
Sov.Phys.Usp. (UFN) {\bf 25} (1982) 195.

\bibitem{Put} P.Putrov, Path integral in energy representation in quantum
mechanics, hep-th/0605169

\bibitem{Nek}
N.Nekrasov, Adv.Theor.Math.Phys. {\bf 7} (2004) 831-864, [arXiv:hep-th/0206161]\\
R.Flume, R.Poghossian, and H.Storch, Mod.Phys.Lett. {\bf A17} (2002)
327-340\\
R.Flume, and R.Poghossian, Int.J.Mod.Phys. {\bf A18} (2003) 2541.

\bibitem{CDG} C.Callan, R.Dashen, and D.Gross, Phys.Rev. {\bf D17}
(1978) 2717.

\bibitem{MMT} A.Mironov, A.Morozov, and T.Tomaras, J.Exp.Theor.Phys.
{\bf 101} (2005) 331-340.

\bibitem{ADHM} M.Atiyah, V.Drinfeld, N.Hitchin, and Yu.Manin,
Phys.Lett. {\bf A65} (1978) 185.

\bibitem{UFN3} A.Morozov,
Phys.Usp.(UFN) {\bf 37} (1994) 1; hep-th/9502091; hep-th/0502010\\
A.Mironov, Int.J.Mod.Phys. {\bf A9} (1994) 4355; Phys.Part.Nucl.
{\bf 33} (2002) 537.

\bibitem{tauf}
A.Gerasimov, S.Khoroshkin, D.Lebedev, A.Mironov, and A.Morozov,
Int.J.Mod.Phys. A10 (1995) 2589-2614\\
A.Mironov, A.Morozov, and L.Vinet, Theor.Math.Phys. {\bf 100} (1995)
890-899 \\
S.Kharchev, A.Mironov, and A.Morozov, Theor.Math.Phys.
{\bf 104} (1995) 129-143\\
A.Mironov, hep-th/9409190; Theor.Math.Phys. {\bf 114} (1998) 127.

\bibitem{quadr} A.Morozov, hep-th/9810031\\
K.Saraikin, hep-th/0604176.

\bibitem{ammMth} A.Alexandrov, A.Mironov, and A.Morozov, to appear

\bibitem{Ko} Kontsevich M.L.,
Funk.Anal.Prilozh. {\bf 25} (1991) v.~2, p.~50 (in Russian)\\
S.Kharchev, A.Marshakov, A.Mironov, A.Morozov, and A.Zabrodin,
Nucl.Phys. {\bf B380} (1992) 181-240; Phys.Lett. {\bf B275} (1992)
311-314\\
S.Kharchev, A.Marshakov, A.Mironov, and A.Morozov, Nucl. Phys. {\bf
B397} (1993) 339; Mod.Phys.Lett. {\bf A8} (1993)
1047-1062; Int.J.Mod.Phys. {\bf A10} (1995) 2015\\
P.Di Francesco, C.Itzykson, and J.-B.Zuber, Commun.Math.Phys. {\bf
151} (1993) 193-219.

\bibitem{coma} Yu.Makkenko, Pis'ma v ZhETF, {\bf 52} (1990)
885-888\\
J.Amborn, J.Jurkievich, and Yu.Makkenko, Phys.Lett. {\bf B251}
(1990) 517\\
T.Morris, Nucl.Phys. {\bf B356} (1991) 703-728\\
A.Anderson, R.C.Meyers, and V.Periwal, Phys.Lett. {\bf B254} (1991)
89-93.

\bibitem{MMMM} Yu.Makeenko, A.Marshakov, A.Mironov, and A.Morozov,
Nucl.Phys. {\bf B356} (1991) 574.

\bibitem{hemo} F.J.Dyson, J.Math.Phys. {\bf 3} (1962) 140-156\\
M.L.Mehta, {\sl Random matrices},
(2nd ed., Academic Press, New York, 1991)\\
E.Br\'ezin, C.Itzykson, G.Parisi, and J.-B.Zuber,
Commun.Math.Phys. {\bf 59} (1978) 35\\
D.Bessis,
Commun.Math.Phys. {\bf 69} (1979) 147\\
D.Bessis, C.Itzykson, and J.B.Zuber,
Adv.Appl.Math. {\bf 1} (1980) 109\\
C.Itzykson, and J.-B.Zuber,
J.Math.Phys. {\bf 21} (1980) 411.

\bibitem{Tch}
A.Gerasimov, A.Marshakov, A.Mironov, A.Morozov, and A.Orlov,
Nucl.Phys. {\bf B357} (1991) 565.

\bibitem{hemophases}
A.Alexandrov, A.Mironov, and A.Morozov, Int.J.Mod.Phys. {\bf A19}
(2004) 4127-4165; Teor.Mat.Fiz. {\bf 142} (2005) 419-488;
Fortsch.Phys. {\bf 53} (2005) 512.

\bibitem{DV} R.Dijkgraaf, and C.Vafa,
Nucl.Phys. {\bf B644} (2002) 3; Nucl.Phys. {\bf
B644} (2002) 21; [arXiv:hep-th/0208048]\\
L.Chekhov, and A.Mironov, Phys.Lett. {\bf B552}
(2003) 293\\
H.Itoyama, and A.Morozov, Nucl.Phys. {\bf B657} (2003) 53;
Phys.Lett. {\bf B555} (2003) 287; Prog.Theor.Phys. {\bf 109} (2003)
433; Int.J.Mod.Phys. {\bf A18}
(2003) 5889\\
L.Chekhov, A.Marshakov, A.Mironov, and D.Vasiliev, Phys.Lett. {\bf
B562} (2003) 323; Proc. Steklov Inst.Math. {\bf 251} (2005) 254.

\bibitem{ammChops}
A.Alexandrov, A.Mironov, and A.Morozov, Int.J.Mod.Phys. (to appear),
hep-th/0412099.

\bibitem{Ch1}  L.Chekhov,
hep-th/9509001.

\bibitem{Kost}
I.K.Kostov,
hep-th/9907060.

\bibitem{David} G.Bonnet, F.David, and B.Eynard, J.Phys. {\bf A33}
(2000) 6739.

\bibitem{Th} A.Klemm, M.Mari\~no, and S.Theisen, JHEP {\bf 0303} (2003) 051.

\bibitem{Giv} A.Givental, Semisimple Frobenius structures at higher genus,
math.AG/0008067.

\bibitem{umo} V.Periwal, and D.Shevitz,
Phys. Rev. Lett. {\bf 64} (1990) 1326; Nucl. Phys. {\bf B344} (1990)
731\\ M.Bowick, A.Morozov, and D.Shewitz, Nucl.Phys., {\bf
B354} (1991) 496-530\\
S.Kharchev, and A.Mironov, Int.J.Mod.Phys. {\bf A7} (1992) 4803-4824\\
A.Mironov, A.Morozov, and G.Semenoff, Int.J.Mod.Phys., {\bf A10}
(1995) 2015.

\bibitem{Mth} P.Horava, and E.Witten, Nucl.Phys. {\bf B460} (1996)
506-524.

\bibitem{landsc} M.R.Douglas, JHEP {\bf 0305} (2003) 046.

\bibitem{GLM} A.Gerasimov, D.Lebedev, and A.Morozov, Int.J.Mod.Phys.
{\bf A6} (1991) 977-988.

\bibitem{BS} T.Banks, W.Fischler, S.H.Shenker, and L.Susskind, Phys.Rev.
{\bf D55} (1997) 5112-5128.


\bibitem{Moyal} H.Weyl, Z.Phys. {\bf 46} (1927) 1; {\sl The Theory
of Groups and Quantum Mechanics}, Dover Publications, New York Inc.
(1931)\\
J.E.Moyal, Proc. Cambridge Phil.Soc. {\bf 45} (1949) 99.

\bibitem{BV} I.A.Batalin, and G.A.Vilkovisky,
  Phys.Lett. {\bf B102} (1981) 27;
  Phys.Rev. {\bf D28} (1983) 2567
  [Erratum-ibid. {\bf D30} (1984) 508]\\
I.A.Batalin, and E.S.Fradkin, Phys.Lett. {\bf B122} (1983) 157\\
B.L.Voronov, and I.V.Tyutin,
  Theor.Math.Phys. {\bf 50} (1982) 218\\
A.~Schwarz, Commun.Math.Phys. {\bf 155} (1993) 249; ibid. {\bf 158}
(1993) 373\\
  A.Losev, and A.Gorodencev, Dombay Seminars on Berkovits Theory,
  Dombay, 2003\\
  A.Losev and D.Krotov, hep-th/0603201.

\bibitem{KM} S.Kharchev, and A.Marshakov, Int.J.Mod.Phys.
{\bf A10} (1995) 1219-1236\\
A.Mironov, On GKM description of multi-criticality in 2d gravity,
Preprint FIAN/TD/16-92.

\bibitem{ACKM} J.Ambjorn, L.Chekhov, and Yu.Makeenko, Phys.Lett.
{\bf B282} (1992) 341-348\\
J.Ambjorn, L.Chekhov, C.F.Kristjansen, and Yu.Makeenko, Nucl.Phys.
{\bf B404} (1993) 127-172; Erratum-ibid. {\bf B449} (1995) 681.

\bibitem{Mir} A.Mironov, Theor.Math.Phys. {\bf 146} (2005) 63.

\bibitem{BE} H.Bateman, {\sl Higher Transcendental Functions}, Volume
II, Mc Graw-Hill, 1953.

\bibitem{KN} I.M.Krichever, and S.P.Novikov,
  Funct.Anal.Appl.  {\bf 21} (1987) 126;
 Funct.Anal.Appl.  {\bf 21} (1987)  294-307 (1987); Funct.Anal.Appl.
{\bf 23} (1989) 19-33\\
R.Dick, DESY-89-160; Lett.Math.Phys. {\bf 18} (1989) 255\\
 M.Schlichenmaier,
  Lett.Math.Phys. {\bf 19} (1990) 151; Lett.Math.Phys. {\bf 19}
  (1990) 327; Lett.Math.Phys. {\bf 26} (1992) 23, hep-th/9207088\\
A.Beilinson, and V.Schechtman, Commun.Math.Phys. {\bf 118} (1988)
651-701\\
A.Morozov, Phys.Lett. {\bf B196} (1987) 325.

\bibitem{Suga} V.G.Knizhnik, and A.B.Zamolodchikov, Nucl.Phys.
{\bf B247} (1984) 83-103.


\bibitem{SW} N.Seiberg, and E.Witten, Nucl.Phys. {\bf B426} (1994)
19.

\bibitem{GKMMM} A.Gorsky, I.Krichever, A.Marshakov, A.Mironov, and A.Morozov,
Phys.Lett. {\bf B355} (1995) 466-477\\
A.Marshakov, {\em Seiberg-Witten Theory and Integrable Systems},
(World Scientific, Singapore, 1999)\\
H.W.Braden and I.Krichever (Eds.), {\em Integrability: The
Seiberg-Witten and Whitham Equations},
(Gordon and Breach, 2000)\\
A.Gorsky and A.Mironov, Integrable Many-Body Systems and Gauge
Theories, hep-th/0011197.

\bibitem{SFT} E.Witten, Nucl.Phys. {\bf B268} (1986) 253\\
B.Zwiebach, Phys.Lett. {\bf B256} (1991) 22-29; Mod.Phys.Lett. {\bf
A7} (1992) 1079-1090; Annals Phys. {\bf 267} (1998) 193-248;
hep-th/9305026.

\bibitem{bi} E.Witten, {\bf B276} (1986) 291; hep-th/9306122\\
A.Sen, and B.Zwiebach, Nucl.Phys. {\bf B414} (1994) 649-714; {\bf
B423} (1994) 580-630; Commun.Math.Phys. {\bf 177} (1996) 305-326\\
 A.Gerasimov, and S.Shatashvili, JHEP {\bf 0106} (2001) 066.

\bibitem{Ey} B.Eynard, JHEP {\bf 0411} (2004) 031\\
 L.Chekhov, and B.Eynard,
hep-th/0504116.

\bibitem{ammEy} A.Alexandrov, A.Mironov, and A.Morozov,
{\it unpublished}

\end{thebibliography}
\end{document}